\documentclass[12pt]{article}
\usepackage{amssymb,amsmath}
\usepackage[noblocks]{authblk}
\usepackage[top=0.75in, bottom=0.75in, left=0.75in, right=0.75in, dvips]{geometry}
\usepackage{caption}
\begin{document}
\textwidth 10.0in 
\textheight 9.0in 
\topmargin -0.60in
\title{Symplectic Analysis of the Two Dimensional Palatini Action}
\author[1,2]{D.G.C. McKeon}
\affil[1] {Department of Applied Mathematics, The
University of Western Ontario, London, ON N6A 5B7, Canada} 
\affil[2] {Department of Mathematics and
Computer Science, Algoma University, \newline Sault St.Marie, ON P6A
2G4, Canada}
\date{}
\maketitle    

\maketitle
\noindent
email: dgmckeo2@uwo.ca\\
PACS No.: \\

\begin{abstract}
The symplectic analysis, initiated by Faddeev and Jackiw, is applied to the first order (Palatini) form of the Einstein-Hilbert action in $1 + 1$ dimensions.  The constraints that arise are shown to result in the same gauge transformations that follow from the first class constraints occurring when the Dirac constraint formalism is applied to this action. Problems associated with gauge fixing are discussed.
\end{abstract}

\section{Introduction}

The first order Einstein-Hilbert(1EH) action (which actually was first discussed by Einstein, not Palatini [1]) is equivalent to the usual second order (2EH) action, both classically [1,2] and quantum mechanically [3] in $d > 2$ dimensions; this is not true when $d = 2$ [4,5].

When $d = 2$, the 1EH action has a number of interesting features.  Although it possesses manifest general covariance, this is not the invariance that results from the first class constraints that are present; the gauge transformation involves a symmetric tensor $\theta_{\mu\nu}$ [6].

A second property of this model is that although it is not a topological theory, it is totally constrained--there are no net physical degrees of freedom once the constraints are taken into account [6].  (Other non-topological models have this property as well [7].)  When path integral quantization is used in conjunction with the Faddeev-Popov (FP) method of treating gauge symmetries [8], it can be shown that all radiative effects vanish [9].  It is not clear if the absence of radiative effects is a direct consequence of there being no net physical degrees of freedom.

In order to gain more insight into these properties of the 1EH action, we apply a different approach to analyzing its canonical structure.  Faddeev and Jackiw [10-12] initiated the so-called ``symplectic'' approach in which a first order form of the Lagrangian is used to analyze its structure.  This approach has been developed so as to uncover the constraints present and to see what gauge transformations they imply [13-15].  (Other topics that have been considered within this Faddeev-Jackiw (FJ) approach are the treatment of Grassmann variables [16,17] and its use in the path integral [18,19].)

Since the 1EH action in $2d$ is already in first order form, it lends itself to the FJ analysis.  We will show how the FJ approach can be used in conjunction with this model to derive the gauge transformations found in refs. [6] through use of the first class constraints that arise in the Dirac approach.  We then will attempt to use the result of this FJ analysis to impose the covariant gauge condition used in the FP quantization of this model in ref. [9].  It turns out that this gauge condition is not compatible with the FJ approach as it involves the time derivative of fields.  A gauge fixing condition that does not involve the time derivative of fields can be used to derive a path integral in the FJ approach that is consistent with the FP path integral.

In the following section we apply the symplectic analysis to the IEH action in $1 + 1$ dimension.  A brief overview of the symplectic method is given in an appendix.

\section{Faddeev-Jackiw Analysis of the 1EH Action in $1 + 1 d$}

The 1EH action in $d$ dimensions
\begin{equation}
S_d = \int d^dx \sqrt{-g}\; g^{\mu\nu} R_{\mu\nu} (\Gamma)
\end{equation}
can be written [1] ($\rm{metric = diag} (+ + + \ldots, -)$)
\begin{equation}
S_d = \int d^dx \, h^{\mu\nu} \left(G_{\mu\nu ,\lambda}^\lambda + \frac{1}{d-1}
G_{\lambda\mu}^\lambda G_{\sigma\nu}^\sigma -  G_{\sigma\mu}^\lambda G_{\lambda\nu}^\sigma\right)
\end{equation}
where [20]
\begin{subequations}
\begin{align}
h^{\mu\nu} &= \sqrt{-g} \;g^{\mu\nu} \quad \left(\det h^{\mu\nu} = -\left( \sqrt{-g}\right)^{d-2}\right)\\
\intertext{and}
G^\lambda_{\mu\nu} &= \Gamma_{\mu\nu}^\lambda - \frac{1}{2}\left(\delta_\mu^\lambda \Gamma_{\rho\nu}^\rho + \delta_\nu^\lambda \Gamma_{\rho\mu}^\rho\right).
\end{align}
\end{subequations}
If $d = 2$, then $S_2$ becomes
\begin{equation}
S_2 = \int d^2x \left[ \pi_{11} \dot{h}^{11} + \pi_{01} \dot{h}^{01} + \pi_{00} \dot{h}^{00} - \zeta_{11}\phi^{11} - \zeta_{01} \phi^{01} - \zeta_{00} \phi^{00}\right]
\end{equation}
where $\dot{f} = f_{,0}$ and
\begin{subequations}
\begin{align}
(\pi_{11},\; \pi_{01},\; \pi_{00}) &= (-G_{11}^0, \;-2 G_{01}^0, \;-G_{00}^0)\\
(\zeta_{11}, \;\zeta_{01},\; \zeta_{00}) &= (G_{11}^1,\; 2 G_{01}^1,\; G_{00}^1)
\end{align}
\end{subequations}
and
\begin{equation}
\left(\phi^{11},\; \phi^{01},\; \phi^{00}\right) = \left(h^{11}_{,1} + 2h^{01}\pi_{00} + h^{11}\pi_{01}, \; 
h^{01}_{,1} + h^{00} \pi_{00} - h^{11}\pi_{11},\; h_{,1}^{00} - h^{00} \pi_{01} - 2h^{01}\pi_{11}\right).
\end{equation}

A canonical analysis of $S_2$ [6] following Dirac's procedure [21,12] leads to the momenta conjugate to $h^{\alpha\beta}$ being $\pi_{\alpha\beta}$; these are six primary second class constraints.  Similarly the momenta conjugate to $\zeta_{11}$, $\zeta_{01}$ and $\zeta_{00}$ all vanish; these primary first class constraints lead to the secondary first class constraints
\begin{equation}
\phi^{11} = \phi^{01} = \phi^{00} = 0.
\end{equation}
Together, these six second class constraints, the six first class constraints and the six associated gauge conditions eliminate all 18 degrees of freedom $h^{\alpha\beta}$, $\pi_{\alpha\beta}$, $G_{\mu\nu}^\lambda$ from phase space.  Using either the approach of either Castellani [22] or of Henneaux, Teitelboim and Zanelli [23], these first class constraints lead to the gauge transformations
\begin{equation}\tag{8a}
\delta h^{\mu\nu} = - \left( \epsilon^{\mu\rho} h^{\nu\sigma} + \epsilon^{\nu\rho} h^{\mu\sigma} \right) \theta_{\rho\sigma}
\end{equation}
\begin{equation}\tag{8b}
\delta G_{\mu\nu}^\lambda = + \epsilon^{\rho\sigma} \left( G_{\mu\rho}^\lambda \theta_{\nu\sigma} + G_{\nu\rho}^\lambda \theta_{\mu\sigma} \right) + \epsilon^{\lambda\sigma} \theta_{\mu\nu\sigma}
\end{equation}
where $\epsilon^{\mu\nu} = -\epsilon^{\mu\nu}(\epsilon^{01} = 1)$ and $\theta_{\mu\nu} = \theta_{\nu\mu}$ is a symmetric gauge function with three independent components.  In ref. [9], this model was quantized using the FP procedure with the path integral; the gauge condition
\begin{equation}\tag{9}
\epsilon_{\lambda\sigma} G_{\mu\nu}^{\lambda , \sigma} \equiv 0
\end{equation}
was employed.

We now will apply the FJ as outlined in the appendix to this action $S_2$.  The action of eq. (4) is of the form of eq. (A.14) with the identifications
\begin{equation}\tag{10a-c}
x_r \rightarrow (h^{\alpha\beta}, \pi_{\alpha\beta}), \quad \eta_A \rightarrow \zeta_{\alpha\beta}, \quad \Omega^A \rightarrow \phi^{\alpha\beta}.
\end{equation}
Since $f_{ab}$ in eq. (A.19a) is now by eq. (4)
\begin{equation}\tag{11}
f_{ab} = \left(\begin{array}{cc} 0 & 1 \\
-1 & 0 \end{array} \right)
\end{equation}
it follows that the Poisson Bracket of eq. (A.24) is now
\begin{equation}\tag{12}
\left\lbrace \phi^{\alpha\beta}, \phi^{\gamma\delta}\right\rbrace = -
 \frac{\partial\phi^{\alpha\beta}}{\partial h^{\lambda\sigma}}
 \frac{\partial\phi^{\gamma\delta}}{\partial \pi_{\lambda\sigma}} +
  \frac{\partial\phi^{\alpha\beta}}{\partial \pi_{\lambda\sigma}}
 \frac{\partial\phi^{\gamma\delta}}{\partial h_{\lambda\sigma}}.
\end{equation}
However, we know from ref. [6] that
\begin{equation}\tag{13a-c}
\left\lbrace \phi^{11}, \phi^{00} \right\rbrace = 2\phi^{01}, \quad
\left\lbrace \phi^{01}, \phi^{00} \right\rbrace = \phi^{00},\quad
\left\lbrace \phi^{01}, \phi^{11} \right\rbrace = -\phi^{11}
\end{equation}
and so all these constraints $\Omega^A$ of eq. (10c) are first class, and no further constraints arise.  The matrix $M_{AS}$ occurring in eq. (A.19b) is given by 
\begin{equation}\tag{14a}
M_{AS} = \left( M^{(h)}, M^{(\pi)}\right)
\end{equation}
where
\begin{equation}\tag{14b}
M^{(h)} =
\left( \begin{array}{ccc}
\partial\phi^{11}/\partial h^{11} \;\;& \partial\phi^{11}/\partial h^{01} \;\;& 
\partial\phi^{11}/\partial h^{00} \\
\partial\phi^{01}/\partial h^{11} \;\;& \partial\phi^{01}/\partial h^{01} \;\;& 
\partial\phi^{01}/\partial h^{00} \\
\partial\phi^{00}/\partial h^{11} \;\;& \partial\phi^{00}/\partial h^{01} \;\;& 
\partial\phi^{00}/\partial h^{00} \end{array}\right)
\end{equation}
and
\begin{equation}\tag{14c}
M^{(\pi)} = \left( 
\begin{array}{ccc}
\partial\phi^{11}/\partial \pi_{11}\;\; & \partial\phi^{11}/\partial \pi_{01} \;\;& \partial\phi^{11}/\partial \pi_{00} \\
\partial\phi^{01}/\partial \pi_{11}\;\; & \partial\phi^{01}/\partial \pi_{01} \;\;& \partial\phi^{01}/\partial \pi_{00} \\
\partial\phi^{00}/\partial \pi_{11}\;\; & \partial\phi^{00}/\partial \pi_{01} \;\;& \partial\phi^{00}/\partial \pi_{00}
\end{array}
\right).
\end{equation}

With $f_{ab}$ given by eq. (A.8) and $F_{ab}$ by eq. (A.18), it follows that the vector $V_a^A$ occurring in eq. (A.26) is given by 
\begin{equation}\tag{15}
V_a^{A\,T} = \left( 
\frac{\partial \phi^A}{\partial\pi_{11}},\;
\frac{\partial \phi^A}{\partial\pi_{01}},\;
\frac{\partial \phi^A}{\partial\pi_{00}}; \;
-\frac{\partial \phi^A}{\partial h^{11}},\;
-\frac{\partial \phi^A}{\partial h^{01}},\;
-\frac{\partial \phi^A}{\partial h^{00}}; -\delta_{A,a+6} \right).
\end{equation}
Since
\begin{equation}\tag{16}
V_a^A F_{ab} = \left( 0; 0; \left\lbrace \phi^A, \phi^{11}\right\rbrace , 
\left\lbrace \phi^A, \phi^{01}\right\rbrace \left\lbrace \phi^A, \phi^{00}\right\rbrace
\right)
\end{equation}
we see that eq. (A.27) is satisfied on account of eq. (13) and so there are no further constraints.

We can now make use of eq. (A.31) to show, for example, that
\begin{equation}
\delta h^{11} = \frac{\partial\phi^{00}}{\partial\phi_{11}} (-1) \theta_{00} +
2 \frac{\partial\phi^{01}}{\partial\phi_{11}} (-1) \theta_{01} + 
\frac{\partial\phi^{11}}{\partial\phi_{11}} (-1) \theta_{11}\nonumber 
\end{equation}
\begin{equation}
= 0 + 2h^{11} \theta_{01} + 0 \nonumber 
\end{equation}
\begin{equation}\tag{17}
= -2\epsilon^1 \rho \; h^{1\sigma} \theta_{\rho \sigma}
\end{equation}
provided $ \theta_{\rho \sigma} = - \theta_{\sigma\rho }$.  Eq. (17) is consistent with eq. (8a).  Similarly, the other transformations appearing in eq. (8a,b) follow from eqs. (A.31, A.37).

In order to remove the arbitrariness in the time development of degrees of freedom that arises due to the presence of a gauge invariance in the initial action, it is necessary to impose extra ``gauge conditions``, one for each first class constraint that is present.  An example of a constraint of the type $\gamma^A$ which appears in eq. (A.38) would be for our model 
\begin{equation}\tag{18}
\gamma^A = G_{\alpha\beta , 1}^0 = \pi_{A,1}.
\end{equation}
Together, $\Omega^A$ appearing in eq. (10c) and $\gamma^A$ appearing in eq. (18) contribute to $F_{ab}$ in eq. (A.39); these together effectively form a set of ``second class'' constraints and both canonical and path integral quantization can proceed as outlined in the appendix.

If we use the gauge fixing of eq. (9), we then have by eq. (5) 
\begin{equation}\tag{19}
\dot{\zeta}_{\alpha \beta} = \pi_{\alpha \beta , 1}
\end{equation}
which is of the form of eq. (A.42).  As is discussed in the appendix, this does not lend itself to path integral quantization.

\section{Discussion}

In this paper we have shown how the symplectic formalism can be applied to a gauge theory with the non-Abelian gauge symmetry of eq. (8).  This approach can be used to derive this symmetry from the constraints that arise when applying the formalism. Path integral quantization of the model has also been discussed using the symplectic approach.

If instead of just considering the $d = 2$ limit of the action in eq. (2), we were to examine what happens when $d > 2$, a much more involved discussion would occur.  When applying the Dirac constraint analysis [25] to this model, tertiary constraints arise.  We anticipate that the analogue of these constraints would also arise when applying the symplectic approach to this action [26].  This problem is currently being considered.

\section*{Acknowledgements}
Roger Macleod had useful input in this work.

\section*{Appendix}

The Dirac approach [21, 12] to analyzing the structure of gauge theories relies on the Hamiltonian canonical formalism.  More recently, Faddeev and Jackiw [10, 11] used a novel approach to the first order form of gauge theories to discuss their dynamics.  In this appendix we will briefly discuss this ``symplectic approach''.

The first order form of the Lagrangian, the starting point of FJ approach, is 
\begin{equation}\tag{A.1}
L(q_c,\dot{q}_c) = a_a(q_c)\dot{q}_a - V(q_c) \qquad (a = 1 \ldots N)
\end{equation}
from which follows the equations of motion
\begin{equation}\tag{A.2}
f_{ab}(q_c) \dot{q}_b = - \frac{\partial V}{\partial q_a}\;\;
\left( f_{ab} \equiv \frac{\partial a_a}{\partial q_b} - \frac{\partial a_b}{\partial q_a}\right).
\end{equation}
The time evolution of each of the $N$ dynamical degrees of freedom $q_a$ is well defined by eq. (A.2) provided $f_{ab}$ can be inverted since
\begin{equation}\tag{A.3}
\dot{q}_a = - f_{ab}^{-1} \frac{\partial V}{\partial q_b}.
\end{equation}
If $f^{-1}$ does not exist, then $f$ has $M \leq N$ eignenvectors with vanishing eigenvalues $v_a^A$
\begin{equation}\tag{A.4}
f_{ab} v_b^A = 0\quad (A = 1 \ldots M).
\end{equation}
From eqs. (A.2, A.4) it follows that
\begin{equation}\tag{A.5}
\Omega^A \equiv v_a^A \frac{\partial V}{\partial q_a} = 0.
\end{equation}
These quantities $\Omega^A$ are the analogues of the ``primary constraints'' in the Dirac procedure [21, 12].

In the original FJ approach [10, 11] these $M$ primary constraints were employed to immediately eliminate $M$ degrees of freedom from $L$ in eq. (A.1).  This is to be done without distinguishing between ``first'' or ``second'' class constraints as is done in the Dirac procedure [21, 12].  With the remaining $N - M$ field variables $\xi_\alpha , L$ in eq. (A.1) takes the form 
\begin{equation}\tag{A.6}
L\left( \xi_\beta , \dot{\xi}_\beta \right) = \overline{a}_\alpha (\xi_\beta) \dot{\xi}_\alpha - V(\xi_\beta).
\end{equation}
it is possible that at this stage the procedure must be repeated to eliminate some of the $\xi_\alpha$.

In ref. [10, 11] it is observed that by the Darboux theorem, a transformation to new variables $\zeta_\alpha(\xi)_\beta$ can be found such that eq. (A.6) becomes 
\begin{equation}\tag{A.7}
L\left( \zeta_\beta , \dot{\zeta}_\beta \right) = \frac{1}{2}\zeta_\alpha w_{\alpha\beta}  \dot{\zeta}_\beta - V(\zeta_\beta).
\end{equation}
where $w_{\alpha\beta}$ is an antisymmetric matrix.  (We shall assume that $\overline{f}_{\alpha\beta} = \partial_\beta\overline{a}_\alpha - \partial_\alpha \overline{a}_\beta$ is invertible.)  If $(N-M)$ is even, $w_{\alpha\beta}$ can be put into the form
\begin{equation}\tag{A.8}
w_{\alpha\beta} = \left( \begin{array}{cc}
0 & 1 \\
-1 & 0 \end{array} \right).
\end{equation}

A method for finding the transformation from eq. (A.1) to eq. (A.6) is given in ref. [11].  This ``symplectic form'' of $L$ is given in eq. (A.7).

From the equations of motion that follow from eqs. (A.6)
\begin{equation}\tag{A.9}
\overline{f}_{\alpha\beta} \dot{\xi}_\beta \equiv \left( \frac{\partial \overline{a}_\alpha}{\partial \xi_\beta} - \frac{\partial \overline{a}_\beta}{\partial \xi_\alpha}\right) \dot{\xi}_\beta = - \frac{\partial V}{\partial\xi_\alpha}
\end{equation}
and (A.7)
\begin{equation}\tag{A.10}
w_{\alpha\beta} \dot{\zeta}_\beta = \frac{\partial V}{\partial\zeta_\beta}.
\end{equation}
It follows that
\begin{equation}\tag{A.11}
w_{\alpha\beta} = \frac{\partial \xi_\rho}{\partial\zeta_\alpha} \overline{f}_{\rho\sigma} \frac{\partial \xi_\sigma}{\partial\zeta_\beta}.
\end{equation}

Using the coordinates $\zeta_\alpha$, the associated canonical variables are
\begin{equation}\tag{A.12}
Q_\alpha = \zeta_\alpha, \quad I\!\!P_\alpha = \frac{1}{2} w_{\alpha\beta} \zeta_\beta .
\end{equation}
The measure for the path integral used to quantize this system is then
\begin{align}
d I\!\!P_\alpha d Q_\beta &= d\zeta_\alpha \nonumber \\
&= d\xi_\alpha \det \left( \frac{\partial \zeta_\rho}{\partial\xi_\sigma}\right)\nonumber
\end{align}
which by eq. (A.11) is
\begin{equation}\tag{A.13}
\hspace{1.4cm}= d\xi_\alpha \det^{\frac{1}{2}} \left(\overline{f}_{\rho\sigma}\right).
\end{equation}
This measure appears in refs. [11, 18, 19].

An alternative to simply using the constraints $\Omega^A$ in eq. (A.5) to eliminate non physical variables is given in ref. [13, 14, 15].  In this approach it is assumed that the $q_a$ appearing in eq. (A.1) can be decomposed into $\eta_A (A = 1 \ldots M)$ and $x_r(r = 1 \ldots N - M)$ so that
\begin{equation}\tag{A.14}
L = a_r (x_s)\dot{x}_r - \eta_A \Omega^A (x_s) - V(x_s)
\end{equation}
with $f_{rs}$ invertible.  The equations of motion following from eq. (A.14) take the form
\begin{equation}\tag{A.15a}
f_{rs} (x_t) \dot{x}_s = - \frac{\partial V}{\partial x_s}
\end{equation}
\begin{equation}\tag{A.15c}
\hspace{1.1cm}0 = \Omega^A.
\end{equation}
The vectors $v_b^A$ now are given by
\begin{equation}\tag{A.16}
v_b^{AT} = \left( 0; \delta_b^A\right)\quad (A = 1 \ldots M,\; b = N - M + 1 \ldots N)
\end{equation}
so that the $\Omega^A$ in eqs. (A.5) are in fact the same as the $\Omega^A$ in eq. (A.14).

In refs. [13, 14, 15] the requirement that these constraints be satisfied at all times is satisfied by supplementing $L$ in eq. (A.14) with a term $\dot{\lambda}_A\Omega^A$   where $\dot{\lambda}_A$ is a Lagrange multiplier.  We can absorb $b - \eta_A$ into $\dot{\lambda}_A$ leaving us with
\begin{equation}\tag{A.17}
L = \left(x_s, \lambda_A \right) = a_r(x_s)\dot{x}_r + \Omega^A (x_s) \dot{\lambda}_A - V(x_s).
\end{equation}
The matrix $f_{ab}$ in eq. (A.2) now is replaced by
\begin{equation}\tag{A.18}
F_{ab} = \left( \begin{array}{cc}
f_{rs} & -M_{rB}^T\\
M_{As} & 0
\end{array} \right)
\end{equation}
where
\begin{equation}\tag{A.19a}
f_{rs} = \frac{\partial a_r}{\partial x_s} - \frac{\partial a_s}{\partial x_r}
\end{equation}
and
\begin{equation}\tag{A19b}
M_{As} = \frac{\partial \Omega^A}{\partial x_s}.
\end{equation}
The standard matrix identities
\begin{equation}\tag{A.20a,b}
M \equiv 
\left( \begin{array}{cc}
A & B\\
C & D
\end{array} \right) =
\left( \begin{array}{cc}
1 & B\\
0 & D
\end{array} \right)
\left( \begin{array}{cc}
\Delta_1 & 0\\
D^{-1}C & 1
\end{array} \right) = 
\left( \begin{array}{cc}
A &0\\
C & 1
\end{array} \right)
\left( \begin{array}{cc}
1 & A^{-1}B\\
0 & \Delta_2
\end{array} \right)
\end{equation}
(where $\Delta_1 = A - BD^{-1}C$ and $\Delta_2 = D - CA^{-1}B$) can be used to show that 
\begin{equation}\tag{A.21a,b}
\det M = \det D \det \Delta_1 = \det A \det \Delta_2
\end{equation}
and
\begin{equation}\tag{A.22a}
M^{-1} = 
\left( \begin{array}{cc}
\Delta_1^{-1} & -\Delta_1^{-1} BD^{-1}\\
-D^{-1}C\Delta_1^{-1} & D^{-1} + D^{-1}C \Delta_1^{-1} BD^{-1}
\end{array} \right)
\end{equation}

\begin{equation}\tag{A.22b}
\hspace{1cm} = 
\left( \begin{array}{cc}
A^{-1} + A^{-1} B\Delta_2^{-1}CA^{-1} & -A^{-1} B \Delta_2^{-1} \\
-\Delta_2^{-1} CA^{-1} & \Delta_2^{-1} 
\end{array} \right).
\end{equation}
By eq. (A.21b) we see from eq. (A.18) that
\begin{equation}\tag{A.23}
\det F_{ab} = \det f_{rs} \det \left( 0 + \frac{\partial\Omega^A}{\partial x_r} f_{rs}^{-1} \frac{\partial\Omega^B}{\partial x_s}\right).
\end{equation}
By construction, $\det f_{rs} \neq 0$.  If the ``Poisson Bracket'' of $\Omega^A$, $\Omega^B$
\begin{equation}\tag{A.24}
\left\lbrace \Omega^A, \Omega^B \right\rbrace \equiv \frac{\partial\Omega^A}{\partial x^r} f_{rs}^{-1} \frac{\partial\Omega^B}{\partial x^s}
\end{equation}
has a non-vanishing determinant when $\Omega^A = 0$, then the equation of motion that follows from eq. (A.17)
\begin{equation}\tag{A.25}
F_{ab} \dot{X}_b = - \frac{\partial V}{\partial X_a}\qquad (X_a \equiv (x_r, \lambda_A))
\end{equation}
dictates the time evolution of $X_a$ as $F_{ab}$ can be inverted.  In this case, the constraints $\Omega^A$ are said to be ``second class'' and can be used to eliminate $M$ of the variables $x_a$ from eq. (A.17).  This reduced set of variables leads to a Lagrangian that can be treated using the original FJ approach of refs. [10, 11] if no further constraints arise following this elimination [26].

If $\left\lbrace \Omega^A, \Omega^B \right\rbrace$ has a vanishing determinant when $\Omega^A = 0$, the $a$ subset of the $\Omega^A$ has a weakly vanishing Poisson Bracket with all other constraints\footnote{A quantity $A$ is ``weakly vanishing'' if it equals zero when the constraints themselves vanish [21].  This is written $A \approx 0$.}. These constraints are called ``first class''; the remaining ones are ``second class''.  If we assume that no second class constraints occur in $L$ of eq. (A.17) (or they have been removed) then $F_{ab}$ has an eigenvector with vanishing eigenvalue
\begin{equation}\tag{A.26}
V_r^A = \left( \begin{array}{cc}
\frac{\partial \Omega^A}{\partial x_t} & f_{tr}^{-1} \\
-\delta_r^A & 
\end{array}\right)
\end{equation}
which by eq. (A.24) satisfies
\begin{equation}\tag{A.27}
V^T F \approx 0
\end{equation}
with an obvious choice for $\delta$.

We note that if $\Omega^A$ is a second class constraint, then by eq. (A.22b) we find from eq. (A.18)
\begin{equation}\tag{A.28}
F^{-1}_{ab} = \left(
\begin{array}{cc}
f^{-1}_{ab} - f_{ac}^{-1} \frac{\partial\Omega^C}{\partial x^c} \Delta_2^{-1CD} \frac{\partial\Omega^D}{\partial x^d}f_{db}^{-1} & f^{-1}_{ac} \frac{\partial\Omega^C}{\partial x^c} \Delta_2^{-1CB} \\
\\
-\Delta_2^{-1AC} \frac{\partial\Omega^C}{\partial x^c} f_{cb}^{-1} & \Delta_2^{-1AB}
\end{array}\right)
\end{equation}
where
\begin{equation}
\Delta_2^{-1} = \frac{\partial\Omega^A}{\partial x^a}f_{ab}^{-1}  \frac{\partial\Omega^B}{\partial x^b} = \left\lbrace \Omega^A, \Omega^B\right\rbrace .\nonumber
\end{equation}
From eqs. (A.25) and (A.28), we see that if a quantity $Z(x_a)$ depends solely on $x_a$, then\newpage
\begin{equation}
\hspace{-7.9cm}\frac{dZ(x_a)}{dt} = \frac{\partial Z}{\partial x_a} \dot{x}_a \nonumber 
\end{equation}
\begin{equation}
= \frac{\partial Z}{\partial x_a} \left(f_{ab}^{-1} - f_{ac}^{-1} \frac{\partial\Omega^C}{\partial x^c} \Delta_2^{-1CD} \frac{\partial\Omega^D}{\partial x^d} f_{db}^{-1} \right) \frac{\partial V}{\partial x_b} \nonumber 
\end{equation}
\begin{equation}\tag{A.29}
\hspace{-1.4cm}= \left\lbrace Z, V \right\rbrace - \left\lbrace Z, \Omega^C\right\rbrace \Delta_2^{-1CD} \left\lbrace \Omega^D, V \right\rbrace .
\end{equation}
Eq. (A.29) defines the ``Dirac Bracket'' [21, 12] of $Z$ and $V, \left\lbrace Z, V\right\rbrace^*$. 

We note that when secondary constraints are present, it may be [27, 28] that the modified symplectic procedure of ref. [13, 14, 15] is not equivalent to the original procedure [10, 11].  We now use these eigenvectors $V$ in the same way that the eigenvector $v$ of eq. (A.4) was used to find the ``primary`` constraints $\Omega^A$; the analogue of eq. (A.5) may reveal further ``secondary'' constraints $\overline{\Omega}^A$.  The whole procedure continues until no further constraints are found [26].

As in the Dirac approach, the presence of first class constraints results in the occurrence of gauge transformations that leave the action invariant.  There are two approaches to finding these gauge transformations from first class constraints in the Dirac approach [22, 23].  The symplectic approach can also be used to uncover gauge transformations [14, 15].  To do this, we start by considering a variation of the action\footnote{Extending this discussion to cover situations when there are secondary or second class constraints is straight forward.} associated with $L$ in eq. (A.17)
\begin{equation}\tag{A.30}
\delta S = \int dt \left[ \left( -f_{rs} (x_s) \dot{x}_s + \frac{\partial\Omega^A}{\partial x_r} \dot{\lambda}_A - \frac{\partial V}{\partial x_r}\right) \delta x_r + \Omega^A (x_s) \delta \dot{\lambda}_A \right].
\end{equation}
We now take $\delta x_r$ to be given by
\begin{equation}\tag{A.31}
\delta x_r = \left( \frac{\partial \Omega^A}{\partial x^t} f_{t\;\,r}^{-1} \right) \theta^A,
\end{equation}
where $\theta^A$is a gauge function.

If there are only primary first class constraints $\Omega^{A(2)}$, then by eq. (24) 
\begin{equation}\tag{A.32}
\frac{\partial \Omega^A}{\partial x^r}
 f_{r\,s}^{-1}\frac{\partial \Omega^B}{\partial x^s}  = \left\lbrace \Omega^A, \Omega^B\right\rbrace \equiv C^{ABC} \Omega^C
\end{equation}
and also
\begin{equation}\tag{A.33}
 f_{rs} \dot{x}_s \left( \frac{\partial \Omega^A}{\partial x^t} 
 f_{tr}^{-1} \theta^A \right) = 
\frac{d \Omega^A}{dt} \theta^A.
\end{equation}
We also know by the equation of motion that follows from eq. (A.17)
\begin{equation}\tag{A.34}
-f_{ab} \dot{x}_b + \frac{\partial \Omega^A}{\Omega x^b} \dot{\lambda}_A - \frac{\partial V}{\partial x^a} = 0
\end{equation}
(which the same as eq. (A.21)) that unless
\begin{equation}
\frac{\partial \Omega^A}{\Omega x^a}f_{ab}^{-1}\frac{\partial V}{\partial x^b} \approx 0 \nonumber
\end{equation}
\begin{equation}\tag{A.35}
\hspace{3.4cm}\equiv U^{AB}\Omega^B
\end{equation}
there will be additional secondary constraints occurring.

Together, by eqs. (A.31, A.32, A.33, A.35) we find that eq. (A.30) becomes
\begin{equation}\tag{A.36}
\delta S = \int dt \left[\left( \Omega^A \dot{\theta}^A - C^{ABC} \Omega^C\theta^B \dot{\lambda}^A - U^{AB} \Omega^B \theta^A \right) + \Omega^A \delta \dot{\lambda}^A\right].
\end{equation}
Reverting to using $-\eta^A$ in place of $\dot{\lambda}^A$ (ie, using eq. (A.14) in place of eq. (A.17)), we see that eq. (A.36) leads to $\delta S = 0$ provided
\begin{equation}\tag{A.37}
\delta \eta^A = \dot{\theta}^A + C^{ABC} \eta^B \theta^C - \theta^B U^{BA}.
\end{equation}
Together, the transformations of eqs. (A.31) and (A.37) leave $s$ following from eq. (A.14) invariant if we only have primary first class constraints $\Omega^A$.

In order to remove the ambiguity on the time development of dynamical variables which follows from the presence of primary first class constraints, extra constraints known as ``gauge conditions'' must be imposed on the theory, one for each primary first class constraint [14, 15].  If these called $\gamma^A(x_s)$, then eq. (A.17) gets replaced by 
\begin{equation}\tag{A.38}
L(x_s, \lambda_A, \kappa_A) = a_r(x_s)\dot{x}_r + \Omega^A (x_s) \dot{\lambda}_A + \gamma^A (x_s) \dot{\kappa}_A - V(x_s)
\end{equation}
and $F_{ab}$ in eq. (A.18) now becomes
\begin{equation}\tag{A.39}
F_{ab} = \left( \begin{array}{ccc}
f_{rs} & -M_{rB}^T & -N^T_{rB} \\
M_{AS} & 0 & 0 \\
N_{AS} & 0 & 0 
\end{array} \right)
\end{equation}
where
\begin{equation}\tag{A.40}
N_{Ar} = \frac{\partial\gamma^A}{\partial x_r}
\end{equation}
so that by (A.21b)
\begin{equation}
\det F_{ab} = \det f_{rs} \det \left[ 0 + \left( \frac{\partial\Omega^A}{\partial x_r}
+ \frac{\partial\gamma^A}{\partial x_r} \right) f_{rs}^{-1}
\left( \frac{\partial\Omega^B}{\partial x_s}
+ \frac{\partial\gamma^B}{\partial x_s} \right) \right].
\end{equation}
Provided $\gamma^A$ has been chosen so that $\det F_{ab} \not\approx 0$, we can proceed as if together $\Omega^A$ and $\gamma^A$ are a set of second class constraints.

Quite often gauge conditions that involve the time derivative of classical variables are used.  (For example, the Lorentz gauge in electrodynamics.)  In this case, the form of the gauge condition usually involves $\dot{\eta}_A$.  If it is of the form 
\begin{equation}\tag{A.42}
\dot{\eta} = \xi_A (x_s)
\end{equation}
then eq. (A.14) becomes
\begin{equation}\tag{A.43}
L = a_r (x_s) \dot{x}_r - \eta_A \Omega^A (x_s) - V(x_s) + \phi_A \left( \dot{\eta}_A - \xi_A (x_s)\right)
\end{equation}
where $\phi_A$ is a Lagrange multiplier and $F_{ab}$ is now
\begin{equation}\tag{A.44}
F_{ab} = \left( \begin{array}{ccc}
f_{ab} & 0 & 0 \\
0 & 0 & 1\\
0 & -1 & 0 \end{array} \right)
\end{equation}
using the basis $q_a = (x_r, \eta_A, \phi_a)$.  Unlike $F_{ab}$ in eq. (A.18) which cannot be inverted if $\Omega^A$ is first class, $F_{ab}$ in eq. (A.44) can be inverted.

When quantizing using the quantum mechanical path integral (QMPI) it has been shown [11, 18, 19] that much like eq. (A.13) the measure acquires a Jacobean factor of $\det^{\frac{1}{2}}F_{ab}$ where $F_{ab}$ with $F_{ab}$ being given by eq. (A.39) when there are primary first class constraints $\Omega^A$ accompanied by gauge conditions $\gamma^A$.  From eq. (A.41) it is apparent that this recovers the usual measure that was obtained by Faddeev using the Dirac approach to constrained systems [24].

However a gauge condition that has the form of eq. (A.42) is not really a ``constraint''; it fixes the time development of $\eta_A$ but does not impose a condition on the dynamical variables $x_s$ in the way that $\gamma^A (x_s) = 0$ does.  Consequently $\det^{\frac{1}{2}} F_{ab}$ with $F_{ab}$ given by eq. (A.44) is not the measure for the QMPI when gauge conditions of the form of eq. (A.42) are considered.  The correct measure when using gauge conditions like (A.42) in conjunction with the symplectic formalism can be obtained by adapting the approach of ref. [29].

Canonical quantization is effected when using the symplectic formalism in the usual way; the Poisson Bracket of eq. (A.24) (or, if there are second class constraints, the Dirac Bracket of eq. (A.29)) is used to define a quantum mechanical commutator.

\end{document}